\documentclass
[superscriptaddress,secnumarabic,amssymb,amsmath,nobibnotes,aps,prd,showkeys,showpacs,nofootinbib,onecolumn]{revtex4}%
\usepackage{graphicx}
\usepackage{epsf}
\usepackage{bm}
\usepackage{amsmath}
\usepackage{amsfonts}
\usepackage{amssymb}%
\providecommand{\U}[1]{\protect\rule{.1in}{.1in}}

\newcommand{\be}{\begin{equation}}
\newcommand{\ee}{\end{equation}}

\newcommand{\mincir}{\raise
-3.truept\hbox{\rlap{\hbox{$\sim$}}\raise4.truept\hbox{$<$}\ }}
\newcommand{\magcir}{\raise
-3.truept\hbox{\rlap{\hbox{$\sim$}}\raise4.truept\hbox{$>$}\ }}

\ifx\pdfoutput\relax\let\pdfoutput=\undefined\fi
\newcount\msipdfoutput
\ifx\pdfoutput\undefined\else
\ifcase\pdfoutput\else
\ifx\paperwidth\undefined\else
\ifdim\paperheight=0pt\relax\else\pdfpageheight\paperheight\fi
\ifdim\paperwidth=0pt\relax\else\pdfpagewidth\paperwidth\fi
\fi\fi\fi
\begin{document}
\title{Inhomogeneous spacetimes in Einstein-\ae ther Cosmology}
\author{Andronikos Paliathanasis}
\email{anpaliat@phys.uoa.gr}
\affiliation{Institute of Systems Science, Durban University of Technology, PO Box 1334,
Durban 4000, South Africa}

\begin{abstract}
We investigate the existence of inhomogeneous Szekeres spacetimes in
Einstein-\ae ther theory. We show that inhomogeneous solutions which can be
seen as extension of the Szekeres solutions existing in Einstein-\ae ther
gravity only for a specific relation between the dimensionless coefficients
which defines the coupling between the \ae ther field with gravity. The two
Szekeres classes of solutions are derived. Also a class of inhomogeneous
FLRW-like spacetimes is allowed by the theory for arbitrary values of the
dimensionless coefficients of the \ae ther field. The stability of the
solutions obtained is performed from where we find that the field equations
evolve more variously in Einstein-\ae ther than in General Relativity, where
isotropic spacetimes and Kantowski-Sachs spacetimes are found to be attractors.

\end{abstract}
\maketitle
\date{\today}

\section{Introduction}

Inhomogeneous spacetimes are of special interest in the gravitational theory,
because in general they are exact solutions of Einstein's General Relativity
(GR) without any symmetries. Inhomogeneous cosmological models are those which
do not satisfy the cosmological principle, but they provide the limit of
Friedmann--Lema\^{\i}tre--Robertson--Walker (FLRW) spacetime \cite{mac01}.
There are many applications of the inhomogeneous spacetimes in cosmological
studies which cover all the different epochs of the universe.\ Indeed,
inhomogeneous universes can be seen as the limit of FLRW spacetimes with
inhomogeneous perturbations, which can describe the CMB anisotropies, as also
the rest of the structure formation \cite{szbook}. As far as the very early
universe is concerned, inhomogeneous spacetimes can have singularities of many
kinds, i.e. isotropic, cigar, pancake, oscillatory and nonscalar. For more
details we refer the reader to the discussion given in \cite{mbook}. \ 

One of the most well-known inhomogeneous spacetimes is the
Lema\^{\i}tre-Tolman-Bondi (LTB) metric which has the spherical symmetry.
\ There are various cosmological applications of the LTB spacetime, and more
specifically as toy models in cosmological studies
\cite{ltb1,ltb2,ltb3,ltb4,ltb5}. LTB spacetimes belong to the more general
family of Szekeres spacetimes \cite{szek}. The latter spacetimes are exact
solutions of GR with an inhomogeneous fluid source where in general the metric
depends on two scale factors. Szekeres spacetimes are categorized in two
classes, the FRLW-like spacetimes where LTB metric belongs.

The recent observation of gravitational waves \cite{ligo1,ligo2} and the
direct observation of the black hole at the center of the galaxy M87 by the
Event Horizon\ Telescope \cite{p1,p2} indicates the validity of GR. However,
in very-large scales, GR is challenged by the cosmological observations
\cite{Teg,Kowal,Komatsu,planck,Ade15}. During the last decades cosmologists
have worked on two main directions on the explanation of the cosmological
observations. The first direction is based on the introduction of an
energy-momentum tensor in Einstein's field equations, where the matter source
is described by an exotic matter source, such that Chaplygin gas,
quintessence, $k-$essence and others \cite{Ratra88,ke,le2,ke3,ke4,ke5,ke6}.
The alternative direction introduced by cosmologists is based on the
introduction of new terms in the Einstein-Hilbert action, the role of these
new terms is to drive the dynamics of the modified field equations such that
to explain the observable phenomena. These kinds of theories are called
alternative/modified theories of gravity \cite{m1,m2,m3,m3a,m4,m5,m6,m7,mod2}.
Therefore, it becomes necessary to study the existence of cosmological
evolution of inhomogeneous spacetimes in these extensions of General relativity.

A family of theories which have drawn the attention of cosmologists are the
Lorentz violated theories. In this work, we are interested in the existence of
inhomogeneous cosmological exact solutions in the Einstein-\ae ther theory
\cite{DJ,DJ2}. In this specific theory, the kinematic quantities of a unitary
time-like vector field coupled to gravity are introduced in the
Einstein-Hilbert action.\ That vector field is called \ae ther and defines a
preferred frame at each point in the spacetime.

Although the gravitational field equations in Einstein-\ae ther theory are of
second-order because of the introduction of the nonlinear terms which follow
by the \ae ther field, there are few known exact solutions in the literature,
some exact cosmological solutions presented recently in \cite{roume}, while
exact solutions which correspond to critical points on the phase space of the
dynamical system are determined in
\cite{ea01,ea02,ea03,ea04,ea05,ea06,ea07,col}. Moreover, exact solutions in
the presence of a scalar field coupled to the \ae ther were found in
\cite{bar01,bar02}.

As far as the inhomogeneous spacetimes n Einstein-\ae ther theory are
concerned, some exact solutions determined in \cite{col}, where the spacetime
admits the spherical symmetry, while cosmological perturbations in
Einstein-\ae ther theory have been studied before йн
\cite{per01,per02}. For spacetimes where they do not admit any isometry, there
are not known exact solutions in the literature. That is specific the problem
that we investigate in this work.

In this study, we focus on the field equations of the Einstein-\ae ther field
for the four-dimensional spacetime which provides the Szekeres spacetimes in
GR. We prove the existence of generalized Szekeres solutions in the context of
Einstein-\ae ther theory. Furthermore, we investigate the general evolution of
the field equations for the case of Szekeres-Szafron spacetimes in
Einstein-\ae ther theory. In particular we write the dynamic equation by using
the 1+3 decomposition and we study the existence of critical points as also
their stability of different values of the Einstein-\ae ther free parameters.
The paper is structured as follows.

In Section \ref{sec2}, we present the field equations in the Einstein-\ae ther
theory. Section \ref{sec3} includes the main analysis of our work where we
solve the field equations of Einstein-\ae ther theory in the context of an
inhomogeneous spacetime which provides the Szekeres family of solutions. The
stability of the exact spacetimes is studied in Section \ref{sec4a}. We found
that as opposed to GR, in Einstein-\ae ther theory the field equations admit
more critical points while the stability of the limits of GR changes such that
spacetimes of special interests to be found as attractors. Finally, in Section
\ref{sec5} we discuss the results and we draw our conclusions.

\section{Einstein-\ae ther gravity}

\label{sec2}

Einstein-\ae ther theory is a Lorentz violated gravitational theory which
consists of GR coupled at second derivative order to a dynamical timelike
unitary vector field, the \ae ther field, $u^{\mu}$. This vector can be
thought as the four-velocity of the preferred frame.

The gravitational Action Integral\ is defined as \cite{jj1}%
\begin{equation}
S_{AE}=\int d^{4}x\sqrt{-g}\frac{R}{2}-\int d^{4}x\sqrt{-g}\left(
K^{\alpha\beta\mu\nu}u_{\mu;\alpha}u_{\nu;\beta}-\lambda\left(  u^{c}%
u_{c}+1\right)  \right)  , \label{ae.01}%
\end{equation}
where $R$ is Ricciscalar of the underlying spacetime with line metric tensor
$g^{\alpha\beta},$ and $K^{\alpha\beta\mu\nu}$ describes the coupling between
the \ae ther field and the gravity, defined as
\begin{equation}
K^{\alpha\beta\mu\nu}\equiv c_{1}g^{\alpha\beta}g^{\mu\nu}+c_{2}g^{\alpha\mu
}g^{\beta\nu}+c_{3}g^{\alpha\nu}g^{\beta\mu}+c_{4}g^{\mu\nu}u^{\alpha}%
u^{\beta}. \label{ae.02}%
\end{equation}

Function $\lambda$ is a Lagrange multiplier which constraints $u^{a}$ to be a
unitary vector field. Parameters $c_{1},~c_{2},~c_{3}$ and $c_{4}$ are
dimensionless constants and define the coupling between the \ae ther field
with gravity. Indeed when constants $c_{Z},~Z=1,2,3,4$ vanish then Action
Integral (\ref{ae.01}) becomes that of the Einstein-Hilbert Action.

The total set field equations follow by variation of the Action Integral
(\ref{ae.01}) with the metric tensor, the \ae ther field and the Lagrange
multiplier $\lambda$. The latter condition, $\frac{\delta S_{AE}}%
{\delta\lambda}=0$ provides the constraint condition%
\begin{equation}
{u^{\mu}}{u_{\mu}+1=0,} \label{ae.03}%
\end{equation}
for the unitarity of the \ae ther field. Variation with respect to the
\ae ther field $\frac{\delta S_{AE}}{\delta u^{\mu}}$, gives the equation of
motion for the vector field $u^{\mu}$ , that is,%
\begin{equation}
c_{4}g^{\mu\nu}u^{\alpha}u_{\nu;\beta}u_{\mu;\alpha}g^{\kappa\beta}%
-c_{4}g^{\mu\kappa}g^{\alpha\lambda}u_{\lambda;\beta}u^{\beta}u_{\mu;\alpha
}-c_{4}g^{\mu\kappa}u^{\alpha}u_{;\beta}^{\beta}u_{\mu;\alpha}-K^{\alpha
\beta\mu\kappa}u_{\mu;\alpha;\beta}-\lambda g^{\alpha\kappa}u_{\alpha}=0.
\label{ae.04}%
\end{equation}

Therefore, the modified gravitational field equations follows by variation
with respect to the metric tensor $\frac{\delta S_{AE}}{\delta g^{\mu\nu}}=0$,
that is,
\begin{equation}
G_{\mu\nu}=R_{\mu\nu}-\frac{1}{2}Rg_{\mu\nu}=T_{\mu\nu}^{\ae }. \label{ae.05}%
\end{equation}

The lhs of the latter expression is the Einstein\ tensor, while the rhs is the
contribution of the \ae ther field in the gravitational field equations which
are presented by the \ae ther energy-momentum tensor~$T_{\mu\nu}^{\ae }$
defined as \cite{col}
\begin{align}
{T_{ab}^{\ae }}  &  =2c_{1}(u_{;a}^{c}u_{c;b}-u_{a;c}u_{b;d}g^{cd})+2\lambda
u_{a}u_{b}+g_{ab}\Phi_{u}\nonumber\\
&  -2[(u_{(a}J^{c}{}_{b)})_{;c}+(u^{c}J_{(ab)})_{;c}-(u_{(a}J_{b)}{}^{c}%
)_{;c}]-2c_{4}\left(  u_{a;c}u^{c}\right)  \left(  u_{b;d}u^{d}\right)  ,
\end{align}
in which ${{J^{a}}_{m}}=-{{K^{ab}}_{mn}u}_{;b}^{n}~,~\Phi_{u}=-K^{ab}{}%
_{cd}u_{;a}^{c}u_{;b}^{d}\,.$

An equivalent way to write the Action Integral (\ref{ae.01}) is with the use
of the kinematic quantities for the \ae ther field $u^{\mu}$. Indeed at the
$1+3$ decomposition quantity $u_{\mu;\nu}$ can be written as
\begin{equation}
u_{\mu;\nu}=\sigma_{\mu\nu}+\omega_{\mu\nu}+\frac{1}{3}\theta h_{\mu\nu
}-\alpha_{\mu}u_{\nu} \label{ae.07}%
\end{equation}
where
\begin{equation}
\alpha_{\mu}=u_{\mu;\nu}u^{\nu}~,~\theta=u_{\mu;\nu}h^{\mu\nu}~,~\sigma
_{\mu\nu}=u_{\left(  \alpha;\beta\right)  }h_{\mu}^{\alpha}h_{\nu}^{\beta
}-\frac{1}{3}\theta h_{\mu\nu}~,~\omega_{\mu\nu}=u_{\left[  \alpha
;\beta\right]  }h_{\mu}^{\alpha}h_{\nu}^{\beta} \label{ae.08}%
\end{equation}
in which $h_{\mu\nu}=g_{\mu\nu}-\frac{1}{u^{\alpha}u_{a}}u_{\mu}u_{\nu}$.

Thus, by using the latter expression the Action Integral (\ref{ae.01}) is
simplified as%

\begin{equation}
S_{EA}=\int\sqrt{-g}dx^{4}\left(  R+c_{\theta}\theta^{2}+c_{\sigma}\sigma
^{2}+c_{\omega}\omega^{2}+c_{\alpha}\alpha^{2}\right)  \label{ae.09}%
\end{equation}
where\ the \ae ther field $u^{\mu}$ has been assumed to be unitary and the the
new coefficient constants are defined as \cite{jj1}
\begin{equation}
c_{\theta}=\frac{1}{3}\left(  3c_{2}+c_{1}+c_{3}\right)  ~,\ c_{\sigma}%
=c_{1}+c_{3}~,\ c_{\omega}=c_{1}-c_{3}~,\ c_{a}=c_{4}-c_{1}, \label{ae.10}%
\end{equation}
and $\sigma^{2}=\sigma^{\mu\nu}\sigma_{\mu\nu}~,~\omega^{2}=\omega^{\mu\nu
}\omega_{\mu\nu}$.

We proceed our analysis by studying the field equations in the case of
inhomogeneous spacetimes.

\section{Szekeres spacetimes}

\label{sec3}

Consider now the inhomogeneous spacetime with line element%
\begin{equation}
ds^{2}=-dt^{2}+e^{2a\left(  t,x,y,z\right)  }dx^{2}+e^{2b\left(
t,x,y,z\right)  }\left(  dy^{2}+dz^{2}\right)  . \label{ae.11}%
\end{equation}
In the context of GR with an inhomogeneous pressureless fluid source with
energy momentum tensor $T_{\mu\nu}=\rho_{m}\left(  t,x,y,z\right)  v_{\mu
}v_{\nu},~$where $v_{\mu}=\delta_{t}^{\mu},$ the line element (\ref{ae.11})
provides the inhomogeneous Szekeres spacetimes. Szekeres spacetimes are exact
solutions of Einstein's GR which lack any symmetry in general, while Szekeres
spacetimes have been characterized as \textquotedblleft
partially\textquotedblright\ localy rotational spacetimes \cite{silent0}.

The magnetic part of the Weyl tensor is zero and since there is not any
pressure component there is no information dissemination with gravitational or
sound waves between the world-lines of neighboring fluid elements, that is why
Szekeres spacetimes belong to the family of silent universes
\cite{silent,silent2}. Furthermore, the rotation and acceleration of the fluid
source are identical zero, while in general the spacetimes are anisotropic
which means that the shear is nonzero as also the expansion rate is nonzero.

In inhomogeneous cosmology, the large-scale structure of the universe is
described by exact solutions, unlike cosmological perturbation theory to
explain the structure formation \cite{szbook}. In addition in \cite{tun11}
proved that small inhomogeneities in the spacetime does not affect necessary
the existence of expansion phases of the universe. Therefore Szekeres
spacetimes can have applications in the description of the universe in the
pre- and after- inflationary epochs. In \cite{per101,per102,per103,per104}
Szekeres spacetimes have been applied as exact perturbations models of an FLRW
background in order to describe the structure formation of the universe. While
in \cite{bbb01,bbb02} the authors proved that a wide class of inhomogeneous
geometries, including the Szekeres spacetimes, can evolve into homogeneous
FLRW geometries for specific initial conditions. A similar result was found in
\cite{bbb03} without imposing an inflationary era in the evolution of the universe.

Szekeres spacetimes are categorized in two subfamilies. Subfamily (A) with
$b_{,x}=0$, describes inhomogeneous spacetimes with two indepedent scale
factors whose time derivatives satisfy the field equations of Kantowski-Sachs
spacetimes. The second subfamily (b) is characterized by the condition
$b_{,x}\neq0$ and corresponds to inhomogeneous FLRW spacetime with only one
free time-dependent scale factor. Szekeres spacetimes have been generalized by
assuming homogeneous fluid source \cite{sz1}, cosmological constant
\cite{sz2}, heat flow \cite{sz3}, electromagnetic field \cite{sz3,sz5},
viscosity\cite{sz6,sz7,sz8} and others \cite{szbook}. Recently exact solutions
for the line element (\ref{ae.11}) with an inflaton have been determined in
\cite{sz9} while some cyclic Szekeres spacetimes were found in \cite{sz10} by
considering the existence of a second phantom ideal gas. We continue by
investigating the existence of exact solutions for the line element
(\ref{ae.11}) in the case of Einstein-\ae ther theory.

For the \ae ther field we do the simplest selection and we assume that it is
the comoving observer $u^{\mu}=\delta_{t}^{\mu}$, which is normalized, i.e.
$u^{\mu}u_{\mu}=-1~$\cite{as1,as2}. For such a selection and for the line
element (\ref{ae.11}) we calculate $\omega=0$ and $\alpha=0$, consequently the
coefficient constants $c_{\omega}$ and $c_{\alpha}$ do not play any role in
the evolution of the dynamical system. For the matter source we consider the
energy momentum tensor $T_{\mu\nu}=\rho_{m}\left(  t,x,y,z\right)  u_{\mu
}u_{\nu}$. The physical reason that we have assumed the \ae ther field to be
the comoving observer is in order the FLRW limit to exists and our solutions
to describe inhomogeneous cosmological solutions.

The energy momentum tensor for the \ae ther field is calculated to be diagonal
with components%
\begin{equation}
T_{t}^{\ae ~t~}=\left(  c_{1}+c_{2}+c_{3}\right)  \left(  a_{,t}\right)
^{2}+2\left(  c_{1}+2c_{2}+c_{3}\right)  \left(  b_{,t}\right)  ^{2}%
+4c_{2}\left(  a_{,t}b_{,t}\right)  , \label{ae.12}%
\end{equation}%
\begin{equation}
T_{x}^{\ae ~x~}=\left(  c_{1}+c_{2}+c_{3}\right)  \left(  2a_{,tt}+\left(
a_{,t}\right)  ^{2}+4a_{,t}b_{t}\right)  +4c_{2}b_{,tt}-2\left(  c_{1}%
-2c_{2}+c_{3}\right)  \left(  b_{,t}\right)  ^{2}, \label{ae.13}%
\end{equation}%
\begin{equation}
T_{y}^{\ae ~y~}=T_{z}^{\ae ~z~}=2c_{2}a_{,tt}-\left(  c_{1}-c_{2}%
+c_{3}\right)  \left(  a_{,t}\right)  ^{2}+2\left(  c_{1}+2c_{2}+c_{3}\right)
\left(  b_{,tt}+\left(  b_{,t}\right)  ^{2}+a_{,t}b_{,t}\right)  ,
\label{ae.14}%
\end{equation}
or equivalently%
\begin{equation}
T_{\mu\nu}^{\ae ~}=\rho^{\text{\ae \ }}u_{\mu}u_{\nu}+p^{\text{\ae }}h_{\mu
\nu}+2q_{(\mu}^{\text{\ae }}u_{\nu)}+\pi_{\mu\nu}^{\text{\ae }} \label{ae.15}%
\end{equation}
in which the physical quantities are defined as \cite{col}
\begin{equation}
\rho^{\text{\ae \ }}=-c_{\theta}\theta^{2}-6c_{\sigma}\sigma^{2}%
~,~p^{\text{\ae \ }}=c_{\theta}\left(  2\theta_{,t}+\theta^{2}\right)
-6c_{\sigma}\sigma^{2} \label{ae.16}%
\end{equation}

\begin{equation}
q_{\mu}^{\text{\ae }}=0~,~\pi_{~x}^{x}=-2\pi_{~y}^{y}=-2\pi_{~z}%
^{z}=-4c_{\sigma}\left(  \sigma_{,t}+\theta\sigma\right)  \label{ae.17}%
\end{equation}
where the shear $\sigma$ and the expansion rate $\theta$ are given by the
following expressions%
\begin{equation}
\theta\left(  t,x,y,z\right)  =a_{,t}+2b_{,t}~,~\sigma\left(  t,x,y,z\right)
=\frac{a_{,t}-b_{,t}}{3}. \label{ae.18a}%
\end{equation}

The equation of motion for the \ae ther field (\ref{ae.04}) provides the
following components%
\begin{equation}
\left(  c_{\theta}+\frac{c_{\sigma}}{6}\right)  \theta_{,x}-3c_{\sigma}%
\sigma_{,x}-9c_{\sigma}\sigma b_{,x}=0, \label{ae.19}%
\end{equation}%
\begin{equation}
\left(  c_{\theta}+\frac{c_{\sigma}}{6}\right)  \theta_{,A}+\frac{3}%
{2}c_{\sigma}\sigma_{,A}+\frac{9}{2}c_{\sigma}\sigma a_{,A}=0, \label{ae.20}%
\end{equation}
where $A,B=y$ or $z$~with $A\neq B$

As we can see for this specific selection of the \ae ther field there are not
any nondiagonal terms at the energy momentum tensor $T_{\mu\nu}^{\ae ~}$ while
the equation of motion for the \ae ther field (\ref{ae.04}) provides
constraints on the space independent variables for the unknown functions
$a\left(  t,x,y,z\right)  $ and $b\left(  t,x,y,z\right)  $.

The diagonal field equations (\ref{ae.05}) are
\begin{align}
0 &  =2\left(  1+2c_{2}\right)  b_{,tt}+2\left(  c_{1}+c_{2}\right)
a_{,tt}-\left(  2c_{1}-4c_{2}-3\right)  \left(  b_{,t}\right)  ^{2}%
+\nonumber\\
&  +\left(  c_{1}+c_{2}\right)  \left(  \left(  a_{,t}\right)  ^{2}%
+4a_{,t}b_{,t}\right)  -4\left(  b_{,\xi\bar{\xi}}\right)  e^{-2b}-\left(
b_{,x}\right)  ^{2}e^{-2a}\label{ae.22}%
\end{align}%
\begin{align}
0 &  =\left(  1+2c_{1}+4c_{2}\right)  b_{,tt}+\left(  1+2\left(  c_{1}%
+c_{2}\right)  \right)  a_{,tt}-\left(  -2c_{1}-4c_{2}-1\right)  \left(
b_{,t}\right)  ^{2}+\left(  1-c_{1}+c_{2}\right)  \left(  a_{,t}\right)
^{2}+\nonumber\\
&  +\left(  1+2c_{1}+4c_{2}\right)  a_{,t}b_{,t}-e^{-2a}\left(  b_{,xx}%
+\left(  b_{,x}\right)  ^{2}-a_{,x}b_{,x}\right)  -2e^{-2b}\left(  a_{,\xi
\bar{\xi}}+a_{,\xi}a_{,\bar{\xi}}\right)  \label{ae.23}%
\end{align}%
\begin{align}
0 &  =\left(  1+c_{1}+3c_{2}\right)  \left(  4b_{,tt}+2a_{,tt}\right)
+\left(  6+4c_{1}+16c_{2}\right)  \left(  b_{,t}\right)  ^{2}+\nonumber\\
&  +\left(  2+4c_{2}\right)  \left(  a_{,t}\right)  ^{2}+4\left(
1+2c_{1}+4c_{2}\right)  b_{,t}a_{,t}-\rho_{m}\left(  t,x,y,z\right)
+\label{ae.24}\\
&  +2e^{-2a}\left(  2a_{x}b_{x}-3b_{x}^{2}-2b_{xx}\right)  -8e^{-2b}\left(
a_{\xi\bar{\xi}}+a_{\xi}a_{\bar{\xi}}+b_{\xi\bar{\xi}}\right)  .\nonumber
\end{align}
where without loss of generality we have assumed $c_{3}=0$ and $\xi=y+iz$.
Specifically, equation (\ref{ae.22}) is the $xx$ component of the field
equations, $G_{x}^{x}=_{eff}T_{~x}^{x}$, equation (\ref{ae.23}) correspond to
the the $\xi\xi$ component $G_{\xi}^{\xi}=_{eff}T_{~\xi}^{\xi},$while equation
(\ref{ae.24}) is $G_{~t}^{t}=_{eff}T_{~t}^{t},$ in which $_{eff}T_{~\mu}^{\mu
}$ is the effective energy momentum tensor $_{eff}T_{~\nu}^{\mu}=T_{\nu
}^{\left(  m\right)  \mu}+T_{\nu}^{\ae ~\mu~}$.

The nondiagonal field equations are the constraint equations presented in
\cite{szek}, they are%
\begin{equation}
b_{,tx}-a_{,t}b_{,x}+b_{,t}b_{,x}=0, \label{ae.25}%
\end{equation}%
\begin{equation}
a_{,tA}+b_{,tA}+a_{,t}a_{,A}-b_{,t}a_{,A}=0, \label{ae.26}%
\end{equation}%
\begin{equation}
b_{,xA}-b_{,x}a_{,A}=0, \label{ae.27}%
\end{equation}%
\begin{equation}
a_{,\bar{\xi}\bar{\xi}}+\left(  a_{,\bar{\xi}}\right)  ^{2}-2a_{\bar{\xi}%
}b_{\bar{\xi}}=0. \label{ae.28}%
\end{equation}
where now, $\xi=y+iz~,~\bar{\xi}=y-iz$ and $A=\xi~$or $\bar{\xi}$.

We continue\ our analysis by assuming the two possible cases (A) $b_{,x}=0$
and (B) $b_{x}\neq0.$

\subsection{Class $A$ with $b_{,x}=0$}

The first class of spacetimes follow by the condition $b_{,x}=0$. Indeed, by
replacing
\begin{equation}
b=\ln\left(  \Phi\left(  t\right)  \right)  +\nu\left(  \xi,\bar{\xi}\right)
~,~a=\ln\left(  R\left(  t,x\right)  +\Phi\left(  t\right)  \mu\left(
x,\xi,\bar{\xi}\right)  \right)  \label{ae.29}%
\end{equation}
in the constraint conditions (\ref{ae.20}) it follows
\begin{equation}
\left(  c_{1}+c_{2}\right)  \mu_{,A}\left(  \Phi R_{,t}-\Phi_{,t}R\right)  =0,
\label{ae.30}%
\end{equation}
while from (\ref{ae.19}) we have$~\xi$
\begin{equation}
\left(  c_{1}+c_{2}\right)  \left(  R_{,tx}\left(  R+\Phi\mu\right)  -\mu\Phi
R_{,x}-\mu_{,x}\left(  \Phi R_{,t}-\Phi_{,t}R\right)  -R_{,t}R_{,x}\right)
=0. \label{ae.31}%
\end{equation}

From the latter conditions we get the subclasses (i)$~c_{1}+c_{2}=0$, (ii)
$R\left(  t,x\right)  =\Phi\left(  t\right)  \omega\left(  x\right)  $ and
(iii) $\mu\left(  x,\xi,\bar{\xi}\right)  =\chi\left(  x\right)  $ ,~$R\left(
t,x\right)  =\Phi^{K}\left(  t\right)  \chi\left(  x\right)  $.

\subsubsection{Subclass $A_{\left(  i\right)  }$}

In the first class where $c_{1}+c_{2}=0,$ by replacing (\ref{ae.29}) in
(\ref{ae.22}) we find
\begin{equation}
\left(  1+2c_{2}\right)  \left(  2\Phi\Phi_{,tt}+\Phi_{,t}^{2}\right)
-4e^{-\nu}\nu_{,\xi\bar{\xi}}=0 \label{ae.32}%
\end{equation}
therefore it follows that $e^{-\nu}\nu_{,\xi\bar{\xi}}=\nu_{0}$, from where we
find that%
\begin{equation}
\nu\left(  \xi,\bar{\xi}\right)  =-2\ln\left(  1+\frac{k}{4}\left(  \xi
-\xi_{0}\right)  \left(  \bar{\xi}-\bar{\xi}_{0}\right)  \right)
\label{ae.36}%
\end{equation}
where without loss of generality we select $\xi_{0}=0$,~$\bar{\xi}_{0}=0$.
Moreover, from (\ref{ae.23}) it follows
\begin{equation}
2\left(  1+2c_{2}\right)  \left(  \Phi R_{,tt}+\Phi_{,t}R_{,t}+\frac{R}{2\Phi
}\left(  \Phi_{,t}\right)  ^{2}\right)  -k\left(  1+4c_{2}\right)  \frac
{R}{\Phi}-\left(  1+2c_{2}\right)  \left(  2e^{-2\nu}\mu_{,\xi\bar{\xi}}%
+k\mu\right)  =0. \label{ae.37}%
\end{equation}
Hence, with the use of the constraint equations (\ref{ae.28}) it follows
\begin{equation}
\mu\left(  x,\xi,\bar{\xi}\right)  =\left(  \frac{U\left(  x\right)  }{2}%
\xi\bar{\xi}+U_{1}\left(  x\right)  \xi+U_{2}\left(  x\right)  \bar{\xi
}+W\left(  x\right)  \right)  e^{\nu\left(  \xi,\bar{\xi}\right)  }
\label{ae.38}%
\end{equation}
in which equations (\ref{ae.32}) and (\ref{ae.36}) are simplified%
\begin{equation}
\left(  1+2c_{2}\right)  \left(  2\Phi\Phi_{,tt}+\Phi_{,t}^{2}\right)  -k=0,
\label{ae.39}%
\end{equation}%
\begin{equation}
2\left(  1+2c_{2}\right)  \left(  \Phi R_{,tt}+\Phi_{,t}R_{,t}+\frac{R}{2\Phi
}\left(  \Phi_{,t}\right)  ^{2}\right)  -k\left(  1+4c_{2}\right)  \frac
{R}{\Phi}-\left(  1+2c_{2}\right)  \left(  2U\left(  x\right)  +kW\left(
x\right)  \right)  =0. \label{ae.40}%
\end{equation}

System (\ref{ae.39}), (\ref{ae.40}) can be easily seen that is integrable.
From equation (\ref{ae.39}) we find that $\Phi\left(  t\right)  $ is expressed
in terms of elliptic integrals, while then equation (\ref{ae.40}) is a linear
equation for $R\left(  t,x\right)  $ in terms of derivatives of $t$, which is
a well-known integrable.

For $k=0$, a closed-form solution can be easily obtained with the use of
power-law exponents as follows
\begin{equation}
\Phi\left(  t\right)  =\Phi_{0}t^{\frac{2}{3}}~,~R\left(  t,x\right)
=\frac{9U\left(  x\right)  }{10\Phi_{0}}t^{\frac{4}{3}}+R_{1}\left(  x\right)
t^{\frac{2}{3}}+R_{2}\left(  x\right)  t^{-\frac{1}{3}}\text{.} \label{ae.41}%
\end{equation}

\subsubsection{Subclass $A_{\left(  ii\right)  }$}

For the second subclass where $R\left(  t,x\right)  =\Phi\left(  t\right)
\omega\left(  x\right)  $, the line element (\ref{ae.11}) becomes
\begin{equation}
ds^{2}=-dt^{2}+\Phi^{2}\left(  t\right)  \left[  \left(  \omega\left(
x\right)  +\mu\left(  x,\xi,\bar{\xi}\right)  \right)  ^{2}dx^{2}%
+e^{2\nu\left(  \xi,\bar{\xi}\right)  }\left(  dy^{2}+dz^{2}\right)  \right]
. \label{ae.42}%
\end{equation}
and by following the same procedure as before we find that~$\nu\left(
\xi,\bar{\xi}\right)  $ and $\mu\left(  x,\xi,\bar{\xi}\right)  $ are given by
the expressions (\ref{ae.36}) and (\ref{ae.38}) while function $\Phi\left(
t\right)  $ satisfies the second-order ordinary differential equation%
\begin{equation}
2\left(  1+c_{1}+3c_{2}\right)  \Phi\Phi_{,tt}+\left(  1+4c_{1}-2c_{2}\right)
\left(  \Phi_{,t}\right)  ^{2}+k=0. \label{ae.43}%
\end{equation}

Furthermore, from (\ref{ae.23}) and with the use of (\ref{ae.43}) the
constraint equation it follows%
\begin{equation}
-4\left(  2U+k\left(  W+\omega\right)  \right)  +\left(  4+k\xi\bar{\xi
}\right)  \left(  \Phi_{,t}\right)  ^{2}\left(  c_{1}+c_{2}\right)  \left(
\xi\bar{\xi}\left(  2U+\omega\right)  +U_{1}\xi+U_{2}\bar{\xi}+4\left(
W+\omega\right)  \right)  =0. \label{ae.44}%
\end{equation}
from where we can infer that all the functions on the parameter $x$ are zero.
Hence, the spacetime\ (\ref{ae.42}) is the homogeneous FLRW spacetime,
consequently from (\ref{ae.24}) it follows that the energy density is
homogeneous. The generic solution of the later system was recently presented
in \cite{roume}.

\subsubsection{Subclass $A_{\left(  iii\right)  }$}

For the third subclass the line element (\ref{ae.11}) is simplified
\begin{equation}
ds^{2}=-dt^{2}+\left(  \Phi^{K}\left(  t\right)  +\Phi\left(  t\right)
\right)  ^{2}\chi^{2}\left(  x\right)  dx^{2}+\Phi^{2}\left(  t\right)
e^{2\nu\left(  \xi,\bar{\xi}\right)  }\left(  dy^{2}+dz^{2}\right)  .
\label{ae.45}%
\end{equation}
where without loss of generality we can select $\chi^{2}\left(  x\right)  =1$.
Function $\nu\left(  \xi,\bar{\xi}\right)  $ is determined by expression
(\ref{ae.36}). However, the two equations (\ref{ae.22}), (\ref{ae.23}) are in
consistency if and only if $K=1,$ from where the latter spacetime reduces to
the homogeneous spacetime (\ref{ae.42}). Hence, there is not any new solution
in that consideration. Before we proceed with the next class of solutions, we
summarize our results in the following statement

For the Szekeres Einstein-\ae ther gravity there exist inhomogeneous solutions
with $b_{,x}$ only when the coefficient constants of the \ae ther field
satisfy the algebraic condition $c_{1}+c_{2}=0.$ Otherwise the spacetime
reduces to the homogeneous FLRW geometry.

\subsection{Class B with $b_{,x}\neq0$}

For the second class it holds $b_{,x}\neq0$, where from the constraint
equations (\ref{ae.25})-(\ref{ae.28}) it follows%
\begin{equation}
a=\ln\left(  h\left(  x\right)  \left(  \Phi_{,x}+\Phi\nu_{,x}\right)
\right)  ~,~b=\ln\left(  R\Phi\right)  +\nu\label{ae.46}%
\end{equation}
in which $R=R\left(  t,x\right)  $ and%
\begin{equation}
\nu\left(  x,\xi,\bar{\xi}\right)  =-\ln\left(  1+\frac{U\left(  x\right)
}{4}\xi\bar{\xi}+\frac{U_{1}\left(  x\right)  }{2}\xi+\frac{U_{2}\left(
x\right)  }{2}\bar{\xi}+W\left(  x\right)  \right)  . \label{ae.47}%
\end{equation}
Hence, there is only one free time dependent function in the spacetime.
Moreover, without loss of generality we can select $h\left(  x\right)  =1$.

We continue by substituting (\ref{ae.46}) in the equations of motion for the
\ae ther field (\ref{ae.19}), (\ref{ae.20}) from where we infer the two
subclasses (i) $c_{1}+c_{2}=0$ and (ii)~$\Phi\left(  t,x\right)  =\Phi\left(
t\right)  \omega\left(  x\right)  $.

\subsubsection{Subclass $B_{\left(  i\right)  }$}

For the first subclass it follows that%
\begin{equation}
2\left(  1+2c_{2}\right)  \Phi\Phi_{,tt}+\left(  1+2c_{2}\right)  \left(
\Phi_{,t}\right)  ^{2}+K\left(  x\right)  =0, \label{ae.48}%
\end{equation}
where%
\begin{equation}
K\left(  x\right)  =U\left(  x\right)  \left(  1+W\left(  x\right)  \right)
-U_{2}\left(  x\right)  U_{1}\left(  x\right)  . \label{ae.49}%
\end{equation}

Equation (\ref{ae.48}) is the modified second Friedmann equation
in\ Einstein-\ae ther theory.

\subsubsection{Subclass $B_{\left(  ii\right)  }$}

For the second subclass where $\Phi\left(  t,x\right)  =\Phi\left(  t\right)
\omega\left(  x\right)  $, from the field equations we find
\begin{equation}
-2\Phi\Phi_{,tt}-\left(  \Phi_{,t}\right)  ^{2}+\left(  \frac{1-K\left(
x\right)  }{\left(  \omega\left(  x\right)  \right)  ^{2}}\right)  =0,
\label{ae.50}%
\end{equation}
from where it follows
\begin{equation}
K\left(  x\right)  =1-k\omega^{2}\left(  x\right)  ,~k=cont\text{. }
\label{ae.51}%
\end{equation}

In this case it is important to mention that the coefficients $c_{1},~c_{2}$
for the \ae ther field do not play any role in the evolution of the dynamical
system. The resulting spacetime is inhomogeneous but the scale factor
$\Phi\left(  t\right)  $ does not depend on the space variable $x$. These
kinds of spacetimes have been determined before in the case of GR with a
homogeneous scalar field \cite{sz9}, or with an isotropic ideal gas
\cite{sz10}.

\section{Dynamical evolution}

\label{sec4a}

In the previous section for simplicity on the presentation of our calculations
we assumed that the matter source is described by a pressureless fluid.
However, if we replace the dust fluid with another ideal gas with constant
equation of parameter $p_{m}=\left(  \gamma-1\right)  \rho_{m}$,~$\gamma
=const$, where the energy momentum tensor for the matter source is $T_{m~\nu
}^{\mu}=\left(  \rho_{m}\left(  t,x,y,z\right)  +p_{m}\left(  t,x,y,z\right)
\right)  v_{\mu}v_{\nu}+p_{m}\left(  t,x,y,z\right)  g_{\mu\nu}$ we get
similar results, that, is we found extensions of the inhomogeneous
Szekeres-Szafron spacetime in Einstein-\ae ther gravity. Recall that as a
Szekeres-Szafron system we refer to the extension of the Szekeres system where
the dust fluid source is replaced by an ideal gas with constant equation of
state parameter \cite{sz1}. Moreover, by assuming a cosmological constant term
in the gravitational Action Integral. Our analysis is still valid and similar
results with that of \cite{sz2} are obtained.

In order to study the stability of the solutions we determined we perform a
detailed analysis of the critical points for the evolution equations. Such
analysis is necessary in order to understand the general evolution of the
spacetime for arbitrary initial conditions.

By using the dynamical quantities{\footnote{{The set of $\left\{  u^{\mu
},e_{\nu}^{\mu}\right\}  $ defines an orthogonal tetrad such that $u_{\mu
}e_{\nu}^{\mu}=0;\ e_{\nu}^{\mu}e_{\mu}^{\lambda}=\delta_{\nu}^{\mu}+u^{\mu
}u_{\nu},$ in order the components of tensors are scalar functions}}} $\rho,$
$p$ and $\pi_{~\nu}^{\text{\ae }\mu}=\pi^{\text{\ae }}e_{\nu}^{\mu},$
kinematic quantities {$\theta$ and }$\sigma$ and the {electric component of
the Weyl tensor, $E_{\nu}^{\mu}=$ $Ee_{\nu}^{\mu}$, the gravitational field
equations are} expressed as a system of first-order algebraic differential
equations%
\begin{align}
\dot{\rho}_{m}+\left(  \rho_{m}+p_{m}\right)  \theta &  =0,\label{sz.01}\\
\dot{\theta}+\frac{\theta^{2}}{3}+6\sigma^{2}+\frac{1}{2}\left(  \rho
_{m}+p_{m}\right)  +\frac{1}{2}\left(  \rho^{\text{\ae \ }}+p^{\text{\ae }%
}\right)   &  =0,\label{sz.02}\\
\dot{\sigma}-\sigma^{2}+\frac{2}{3}\theta\sigma+E+\frac{1}{2}\pi
^{\text{\ae }}  &  =0,\label{sz.03}\\
\dot{E}+\frac{1}{2}\dot{\pi}^{\text{\ae }}+\left(  3\sigma+\theta\right)
\left(  E+\frac{1}{6}\pi^{\text{\ae }}\right)  +\frac{1}{2}\left(  \rho
_{m}+p_{m}\right)  \sigma+\frac{1}{2}\left(  \rho^{\text{\ae \ }%
}+p^{\text{\ae }}\right)  \sigma &  =0, \label{sz.04}%
\end{align}
where the algebraic constraint is
\begin{equation}
\frac{\theta^{2}}{3}-3\sigma^{2}+\frac{R^{\left(  3\right)  }}{2}-\rho
_{m}-\rho^{\text{\ae \ }}=0. \label{sz.05}%
\end{equation}
{where $\dot{}$ denotes the directional derivative along the vector field
$u^{\mu}$, i.e. $\dot{}=u^{\mu}\nabla_{\mu},~$and$~$}$R^{\left(  3\right)  }$
describes the curvature of the three-dimensional hypersurface,. We recall that
for the line element (\ref{ae.11}) the magnetic part of the Weyl tensor and
the vorticity term are identical zero. When $p=\pi^{\text{\ae }}=0$, system
(\ref{sz.01})-(\ref{sz.04}) reduce to the known as Szekeres-Szafron system is
recovered \cite{silent}.

By using expressions (\ref{ae.16}), (\ref{ae.17}) and the equation of state
for the ideal gas, we can write the field equations as a system of four
first-order ordinary differential equations of the form%
\begin{align}
\dot{\theta}  &  =\Theta\left(  \rho_{m},\theta,\sigma,E;\delta\right)
,\label{sz.06}\\
\dot{\rho}_{m}  &  =f_{1}\left(  \rho_{m},\theta,\sigma,E;\delta\right)
,\label{sz.07}\\
\dot{\sigma}  &  =f_{2}\left(  \rho_{m},\theta,\sigma,E;\delta\right)
,\label{sz.08}\\
\dot{E}  &  =f_{3}\left(  \rho_{m},\theta,\sigma,E;\delta\right)  .
\label{sz.09}%
\end{align}
in which $\alpha$ contents the free parameters of our model, i.e.
$\delta=\delta\left(  \gamma,c_{1},c_{2}\right)  $.

We continue by defining the dependent and independent variables
\begin{align}
\rho_{m}  &  =\left(  1+\frac{c_{1}}{3}+c_{2}\right)  \Omega_{m}\left(
\tau\right)  \theta^{2}~,~R^{\left(  3\right)  }=\left(  1+\frac{c_{1}}%
{3}+c_{2}\right)  \Omega_{R}\left(  \tau\right)  \theta^{2}~,~\label{sz.10}\\
\sigma\left(  \tau\right)   &  =\left(  1+c_{1}+3c_{2}\right)  \Sigma\left(
\tau\right)  \theta^{2}~\ \text{and }E=\left(  1+c_{1}+3c_{2}\right)
\epsilon\left(  \tau\right)  \theta^{2}~,~dt=\theta d\tau\label{sz.11}%
\end{align}
the modified Szekeres system (\ref{sz.06})-(\ref{sz.09}) is written as a
system of three first-order ordinary differential equations of the form%
\begin{align}
\Omega_{m}^{\prime}  &  =F_{1}\left(  \Omega_{m},\Sigma,\epsilon
;\delta\right)  ,\label{sz.12}\\
\Sigma^{\prime}  &  =F_{2}\left(  \Omega_{m},\Sigma,\epsilon;\delta\right)
,\label{sz.13}\\
\epsilon^{\prime}  &  =F_{3}\left(  \Omega_{m},\Sigma,\epsilon;\delta\right)
. \label{sz.14}%
\end{align}
where prime \ denotes differentiation with respect to the variable $\tau$ and
functions $F_{1},$ $F_{2}$ and $F_{3}$ are defined as follows
\begin{equation}
F_{1}=\frac{\Omega_{m}}{3}\left(  \left(  2-3\gamma\right)  \left(
1-\Omega_{m}\right)  -36\left(  2\alpha-1\right)  \beta\Sigma^{2}\right)  ,
\label{sz.15}%
\end{equation}%
\begin{equation}
\alpha F_{2}=\epsilon-\frac{\left(  2-\alpha\left(  4+\left(  2-3\gamma
\right)  \Omega_{m}\right)  \right)  }{6}\Sigma-\beta\Sigma^{2}+6\left(
1-2\alpha\right)  \alpha\beta\Sigma^{3}, \label{sz.16}%
\end{equation}%
\begin{align}
\alpha F_{3}  &  =-\frac{\epsilon}{3}\left(  1-\alpha\left(  2\left(
1-\alpha\right)  -\left(  2-3\gamma\right)  \Omega_{m}\right)  \right)
\nonumber\\
&  +\left(  \left(  \frac{\left(  2-\alpha\left(  8+\alpha\left(
\beta-7\right)  \right)  \right)  }{18}+\frac{\alpha\left(  2-3\gamma
-2\alpha\beta\right)  }{18}\Omega_{m}\right)  -\beta\left(  2-\alpha
+2\alpha^{2}\right)  \epsilon\right)  \Sigma+\nonumber\\
&  +\frac{\beta}{3}\left(  3+\alpha\left(  \alpha\left(  4+72\epsilon\right)
-7-36\epsilon\right)  \right)  \Sigma^{2}+\beta\left(  2\beta+\alpha\left(
2-\beta+\alpha\left(  \alpha\beta-4\right)  \right)  \right)  \Sigma^{3}.
\label{sz.18}%
\end{align}
The new parameters $\alpha$ and $\beta$ are defined as $\alpha=1-c_{1}$ and
$\beta=1+c_{1}+3c_{2}$.

Furthermore, equation (\ref{sz.05}) provides the constraint equation%
\begin{equation}
\Omega_{R}=-2\left(  1+9\left(  1+2\alpha\right)  \beta\Sigma^{2}-\Omega
_{m}\right)  . \label{sz.19}%
\end{equation}

We continue by determining the critical points of the dynamical system
(\ref{sz.12})-(\ref{sz.14}) and study the physical properties on the solution
at the critical points as also the stability. In order to compare the results
of Einstein-\ae ther gravity with that of GR, let us proceed with the
stability analysis of the Szekeres-Szafron \cite{silent}.

\subsection{Stability analysis for the Szekeres-Szafron system in GR}%

\begin{table}[tbp] \centering
\caption{Critical points for the Szekeres-Szafron system in General Relativity}%
\begin{tabular}
[c]{cccccc}\hline\hline
\textbf{Point} & $\left(  \mathbf{\Omega,\Sigma,\varepsilon}\right)  $ &
\textbf{Physical} & $^{\left(  3\right)  }\mathbf{R}$ & \textbf{Spacetime} &
\textbf{Stability}\\\hline
$A_{1}$ & $\left(  0,0,0\right)  $ & Yes & $<0$ & FLRW (Milne Universe) &
Unstable\\
$A_{2}$ & $\left(  1,0,0\right)  $ & Yes & $=0$ & FLRW (Spatially Flat) &
Unstable\\
$A_{3}$ & $\left(  0,-\frac{1}{3},0\right)  $ & Yes & \thinspace$=0$ & Bianchi
I (Kasner universe) & Stable\\
$A_{4}$ & $\left(  0,\frac{1}{6},0\right)  $ & Yes & $<0$ & Kantowski-Sachs &
Unstable\\
$A_{5}$ & $\left(  3\left(  1-\gamma\right)  ,\frac{3\gamma-2}{6}%
,\frac{\left(  \gamma-1\right)  \left(  3\gamma-2\right)  }{6}\right)  $ &
No &  &  & \\
$A_{6}$ & $\left(  0,\frac{1}{3},\frac{2}{9}\right)  $ & Yes & $=0$ & Bianchi
I (Kasner universe) & Stable\\
$A_{7}$ & $\left(  0,-\frac{1}{12},\frac{1}{32}\right)  $ & Yes & $<0$ &
Kantowski-Sachs & Unstable\\
$A_{8}$ & $\left(  3\left(  3-4\gamma\right)  ,\frac{2}{3}-\gamma,\frac
{\gamma\left(  3\gamma-2\right)  }{6}\right)  $ & No &  &  & \\\hline\hline
\end{tabular}
\label{pointA}%
\end{table}%

For $c_{1}=c_{2}=0$, the dynamical system (\ref{sz.12})-(\ref{sz.14}) reduces
to that of the Szekeres-Szafron system. Every critical point $P=\left(
\Omega_{m}\left(  P\right)  ,\Sigma\left(  P\right)  ,\epsilon\left(
P\right)  \right)  $ is a solution of the following algebraic system%
\begin{equation}
F_{1}\left(  \Omega_{m},\Sigma,\epsilon;\gamma,0,0\right)  =0~,~F_{2}\left(
\Omega_{m},\Sigma,\epsilon;\gamma,0,0\right)  =0,~F_{3}\left(  \Omega
_{m},\Sigma,\epsilon;\gamma,0,0\right)  =0. \label{sz.20}%
\end{equation}

Point $A_{1}$ with coordinates $\left(  0,0,0\right)  $ describes a FLRW
spacetime with nonzero negative curvature, i.e. $\Omega_{R}=-2$, that means
the solution at point $O$ is that of the Milne universe. The eigenvalues of
the linearized system around the critical point are found to be $e_{1}\left(
A_{1}\right)  =\frac{1}{3}~,~e_{2}\left(  A_{1}\right)  =\frac{1}{3}%
~,~e_{3}\left(  A_{1}\right)  =\frac{2}{3}-\gamma,$from where we can infer
that the solution at the point is always unstable. \qquad

Point $A_{2}=\left(  1,0,0\right)  $ describes a spatially flat FLRW universe
where, the eigenvalues are calculated to be $e_{1}\left(  A_{2}\right)
=1-\frac{\gamma}{2},~e_{2}\left(  A_{2}\right)  =\frac{2}{3}-\gamma
,~e_{3}\left(  A_{2}\right)  =\frac{2}{3}-\gamma.$Hence point $A_{2}$ is a
saddle point.

Point $A_{3}=\left(  0,-\frac{1}{3},0\right)  $ describes a Kasner universe,
while it is an attractor since all the eigenvalues of the linearized system is
always negative, that is, $e_{1}\left(  A_{3}\right)  =-1,~e_{2}\left(
A_{3}\right)  =-2,~e_{3}\left(  A_{3}\right)  =\gamma-2$.

Point $A_{4}$ with coordinates $\left(  0,\frac{1}{6},0\right)  $ describes a
Kantowski-Sachs universe with eigenvalues $e_{1}\left(  A_{4}\right)
=-\frac{1}{2},~e_{2}\left(  A_{4}\right)  =\frac{1}{2}$ and $e_{3}\left(
A_{4}\right)  =\gamma-1$. Point $A_{4}$ is a saddle point.

Point $A_{5}$ with coordinates $\left(  3\left(  1-\gamma\right)
,\frac{3\gamma-2}{6},\frac{\left(  \gamma-1\right)  \left(  3\gamma-2\right)
}{6}\right)  $ is physical only when $\gamma=1$~and reduces to $A_{4}$.

Point $A_{6}=\left(  0,\frac{1}{3},\frac{2}{9}\right)  $ describes a Kasner
universe. The eigenvalues are $e_{1}\left(  A_{6}\right)  =-\frac{2}{3}~$
,~$e_{2}\left(  A_{6}\right)  =-\frac{5}{3}$, $e_{3}\left(  A_{6}\right)
=\gamma-2$, hence it is an attractor.

Point $A_{7}=\left(  0,-\frac{1}{12},\frac{1}{32}\right)  $ describes an
unstable Kantowski-Sachs universe; the eigenvalues are derived to be
$e_{1}\left(  A_{6}\right)  =\frac{5}{8}~$ ,~$e_{2}\left(  A_{6}\right)
=-\frac{1}{4}$, $e_{3}\left(  A_{6}\right)  =\gamma-\frac{3}{4},$ which means
that the solution at point $A_{7}$ is always unstable.

Finally point $A_{8}=\left(  3\left(  3-4\gamma\right)  ,\frac{2}{3}%
-\gamma,\frac{\gamma\left(  3\gamma-2\right)  }{6}\right)  $ is unphysical
because $\Omega_{m}\left(  A_{8}\right)  <0$. Hence we do not study its properties.

\begin{figure}[ptb]
\textbf{ \includegraphics[width=0.5\textwidth]{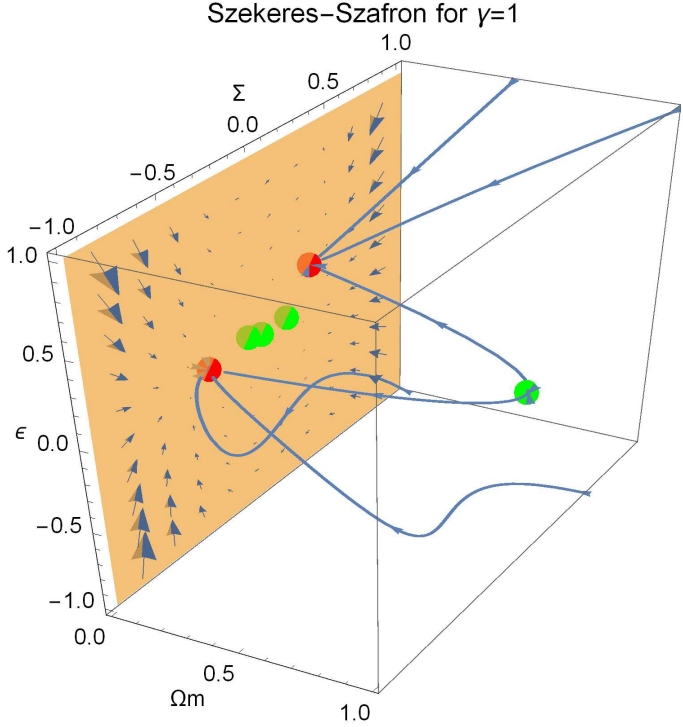} \newline%
}\caption{Phase portrait for the Szekeres-Szafron system in General
Relativity, for $\gamma=1$. With red color are marked the two Kasner
attractors while with green color are marked the unstable critical points.}%
\label{fig1}%
\end{figure}

The above results are collected and presented in table \ref{pointA}. The phase
portrait of the\ Szekeres-Szafron system in presented in Fig. \ref{fig1} where
the critical points are marked.

\subsection{Stability analysis in the Einstein-\ae ther gravity}

We continue by performing the stability analysis for the Szekeres-Szafron
system (\ref{sz.15})-(\ref{sz.10}) in the Einstein-\ae ther theory, i.e.
$c_{1}c_{2}\neq0$. For our analysis we use the constraint condition
$c_{1}+c_{2}=0$ which has been obtained before by the space-constraint
equations. Therefore,by using the latter condition in the dynamical
system(\ref{sz.15})-(\ref{sz.10}) we find the following critical points:

Point $B_{1}=\left(  0,0,0\right)  $ which describes the Milne universe it is
an unstable, the eigenvalues are
\[
e_{1}\left(  B_{1}\right)  =\gamma-\frac{2}{3}~,~e_{2,3}\left(  B_{1}\right)
=\frac{1}{3}\left(  1-c_{1}\right)  \pm\sqrt{c_{1}\left(  c_{1}+1\right)  }.
\]

Point $B_{2}=\left(  1,0,0\right)  $ describes a spatially flat FLRW
spacetime, where the eigenvalues are found to be
\[
e_{1}\left(  B_{2}\right)  =\frac{2}{3}-\gamma~,~e_{2,3}\left(  B_{2}\right)
=\frac{1}{12}\left(  10-9\gamma-4c_{1}\pm\sqrt{16c_{1}^{2}+8c_{1}\left(
3\gamma-4\right)  +\left(  2+3\gamma^{2}\right)  }\right)
\]
from where we can infer that the point describes a stable solution when
\begin{equation}
\left\{  \gamma\in\left(  1,\frac{3+\sqrt{3}}{3}\right)  ,c_{1}>-\frac
{3\gamma^{2}-8\gamma+4}{2\left(  \gamma-1\right)  }\right\}  \cup\left\{
\gamma\in\left(  \frac{3+\sqrt{3}}{3},2\right)  ,-\frac{3\gamma^{2}-8\gamma
+4}{2\left(  \gamma-1\right)  }<c_{1},~c_{1}\neq1\right\}  . \label{con.01}%
\end{equation}

Point $B_{3}=\left(  0,\frac{1}{6c_{1}-3},0\right)  $ describes a Kasner
universe and exists when $c_{1}\neq\frac{1}{2}$. The eigenvalues of the
linearized system are calculated to be
\[
e_{1}\left(  B_{3}\right)  =\gamma-2~,~e_{2,3}\left(  B_{3}\right)  =-\frac
{3}{2}\pm\frac{\sqrt{9+8c_{1}}}{6},
\]
from where we can infer that the point is stable for $-\frac{9}{8}<c_{1}$ with
$c_{1}\neq\frac{1}{2},1$.

Point $B_{4}=\left(  0,\frac{1}{24}\left(  1+\frac{\sqrt{9+4\left(
c_{1}-7\right)  c_{1}}}{1-2c_{1}}\right)  ,\frac{9+c_{1}\left(  43-4c_{1}%
\left(  15+2\left(  c_{1}-7\right)  c_{1}\right)  \right)  +\left(
3+c_{1}-14c_{1}^{2}+4c_{1}^{3}\right)  \sqrt{9+4\left(  c_{1}-7\right)  c_{1}%
}}{576\left(  2c_{1}-1\right)  }\right)  $ exists when $c_{1}\in\left(
-\infty,\frac{7-2\sqrt{10}}{2}\right)  \cup\left(  \frac{7+2\sqrt{10}}%
{2},\infty\right)  $. The eigenvalues of point $B_{4}$ are determined
numerically and the region of the parameters $\gamma$ and $c_{1}$ where the
point $B_{4}$ describes a stable solution for $c_{1}\lesssim-0.65$ and
$c_{1}\gtrsim6.5$ independent from the value of parameter $\gamma.$

\begin{figure}[ptb]
\textbf{ \includegraphics[width=0.4\textwidth]{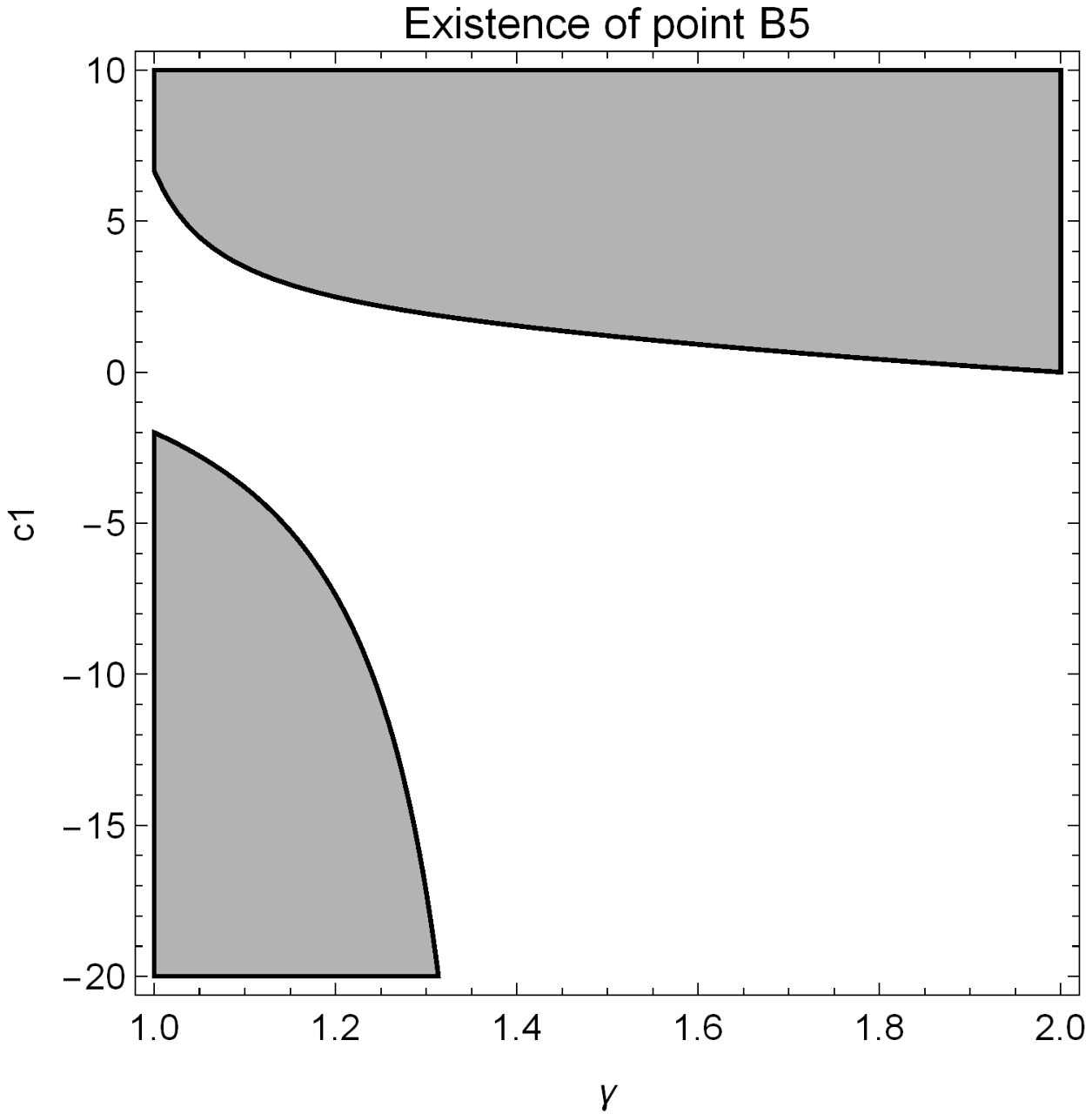} }%
\includegraphics[width=0.4\textwidth]{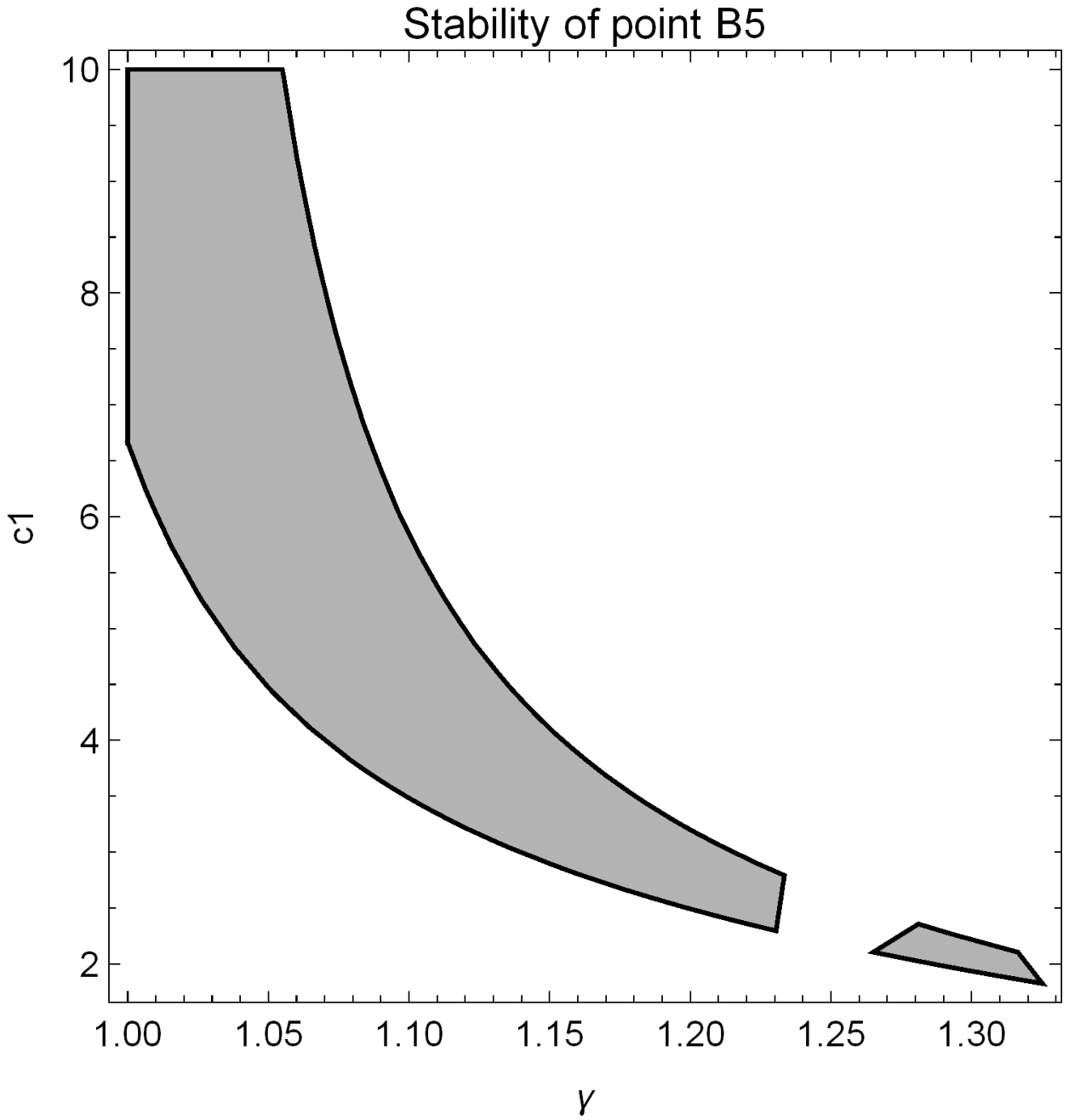}\caption{Region plot for the
free parameter $\gamma$ and $c_{1}$ point $B_{5}$ exists (Left fig.) and
$B_{5}$ is an attractor (Right fig.) }%
\label{fig3}%
\end{figure}

Point $B_{5}=\left(  \Omega_{m}\left(  B_{5}\right)  ,\Sigma\left(
B_{5}\right)  ,\epsilon\left(  B_{5}\right)  \right)  $ with coordinates as
given in Appendix \ref{ap1}. In Einstein-\ae ther theory, point $B_{5}$ exists
for specific values of the free parameters $c_{1}$ and $\gamma$. The region of
the existence of the critical point is presented in Fig. \ref{fig3}. Moreover,
in Fig. \ref{fig3} the region of the free parameters is given where the point
$B_{5}$ describes a stable solution.

Point $B_{6}=\left(  0,\frac{1}{3-6c_{1}},\frac{2}{9-18c_{1}}\right)  $ exists
when $c_{1}\neq\frac{1}{2}$. Point $B_{6}$ describes a Kasner universe. The
eigenvalues of the linearized system are derived to be%
\begin{equation}
e_{1}\left(  B_{6}\right)  =\gamma-2~,~e_{2,3}\left(  B_{6}\right)  =-\frac
{1}{6}\left(  7+4c_{1}\pm\sqrt{9+16c_{1}^{2}}\right)  .
\end{equation}
Hence point $B_{6}$ is a source when $-\frac{5}{7}<c_{1},~c_{1}\neq\frac{1}%
{2},1$.

Point $B_{7}=\left(  0,\frac{2c_{1}-1+\sqrt{9+4c_{1}(c_{1}-7)}}{24\left(
2c_{1}-1\right)  },-\frac{9-43c_{1}+60c_{1}^{2}-56c_{1}^{3}+8c_{1}^{4}+\left(
3+c_{1}-14c_{1}^{2}+4c_{1}^{3}\right)  \sqrt{9+4\left(  c_{1}-7\right)  c_{1}%
}}{576\left(  2c_{2}-1\right)  }\right)  $, which exists for $c_{1}\in\left(
-\infty,\frac{7-2\sqrt{10}}{2}\right)  \cup\left(  \frac{7+2\sqrt{10}}%
{2},\infty\right)  .$ Point $B_{7}$ describes a Kantowski-Sachs universe which
is stable when $c_{1}>\frac{7+2\sqrt{10}}{2}$.

\begin{figure}[ptb]
\textbf{ \includegraphics[width=0.5\textwidth]{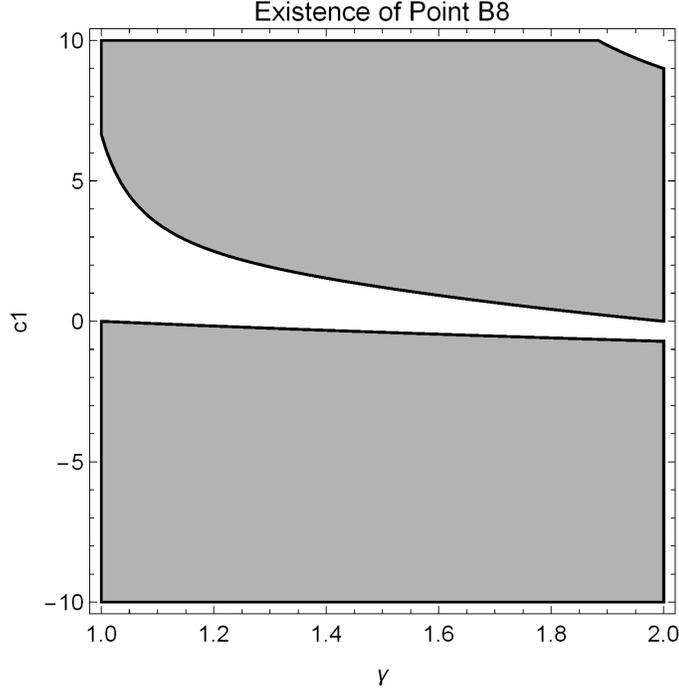}}\caption{Region plot
for the free parameter $\gamma$ and $c_{1}$ point $B_{8}$ it is physical
accepted.}%
\label{fig4}%
\end{figure}

Point $B_{8}$ with coordinates $B_{8}=\left(  \Omega_{m}\left(  B_{8}\right)
,\Sigma\left(  B_{8}\right)  ,\epsilon\left(  B_{8}\right)  \right)  $ as they
are given in Appendix \ref{ap1} exists for the range of variables as they are
given in Fig. \ref{fig4} and describes a Kantowski-Sachs universe. As far as
the stability is concerned from numerical simulations we found that the point
describes an unstable solution. 

In table \ref{pointB} we collect the results of the critical point analysis.%

\begin{table}[tbp] \centering
\caption{Critical points for the Szekeres-Szafron system in Einstein-aether gravity}%
\begin{tabular}
[c]{ccccc}\hline\hline
\textbf{Point} & \textbf{Physical} & $^{\left(  3\right)  }\mathbf{R}$ &
\textbf{Spacetime} & \textbf{Stability}\\\hline
$B_{1}$ & Yes - Always & $<0$ & FLRW (Milne Universe) & Unstable\\
$B_{2}$ & Yes- Always & $=0$ & FLRW (Spatially Flat) & Stable - See Eq.
(\ref{con.01})\\
$B_{3}$ & Yes-$~c_{1}\neq\frac{1}{2}$ & \thinspace$=0$ & Bianchi I (Kasner
universe) & Stable for~$c_{1}>-\frac{9}{8}$\\
$B_{4}$ & Yes-$~\left\vert c_{1}\right\vert >\frac{7-2\sqrt{10}}{2}$ & $<0$ &
Kantowski-Sachs & Stable for $c_{1}\lesssim-0.65$ and $c_{1}\gtrsim6.5$\\
$B_{5}$ & Yes - See Fig. \ref{fig3} &  & Kantowski-Sachs & Stable - See Fig.
\ref{fig3}\\
$B_{6}$ & Yes - $c_{1}\neq\frac{1}{2}$ & $=0$ & Bianchi I (Kasner universe) &
Stable for$\ c_{1}>-\frac{5}{7}$\\
$B_{7}$ & Yes~-$~\left\vert c_{1}\right\vert >\frac{7-2\sqrt{10}}{2}$ & $<0$ &
Kantowski-Sachs & Stable for$~c_{1}>\frac{7+2\sqrt{10}}{2}$.\\
$B_{8}$ & Yes - See Fig. \ref{fig4} &  & Kantowski-Sachs &
Unstable\\\hline\hline
\end{tabular}
\label{pointB}%
\end{table}%

We can see that points $B_{I}$ reduce to points $A_{I}$ when $c_{1}=0$, hence,
the limit of General Relativity is recovered. However, there exit two
additional critical points which describe Kantowski-Sachs universe. Point
$B_{5}$ describes an empty Kantowski-Sachs universe while point $B_{8}$
describes a Kantowski-Sachs universe with matter source.

In addition, the stability of the solutions change. While in GR only the
Kasner universes are attractors that it is not true for the Szekeres-Szafron
system in Einstein-\ae ther theory. For example, the solution at point $A_{2}$
in GR which describes a spatially flat FLRW spacetime dominated by the ideal
gas is always an unstable point, while in Einstein-\ae ther the point can be
an attractor.

\begin{figure}[ptb]
\textbf{ \includegraphics[width=0.5\textwidth]{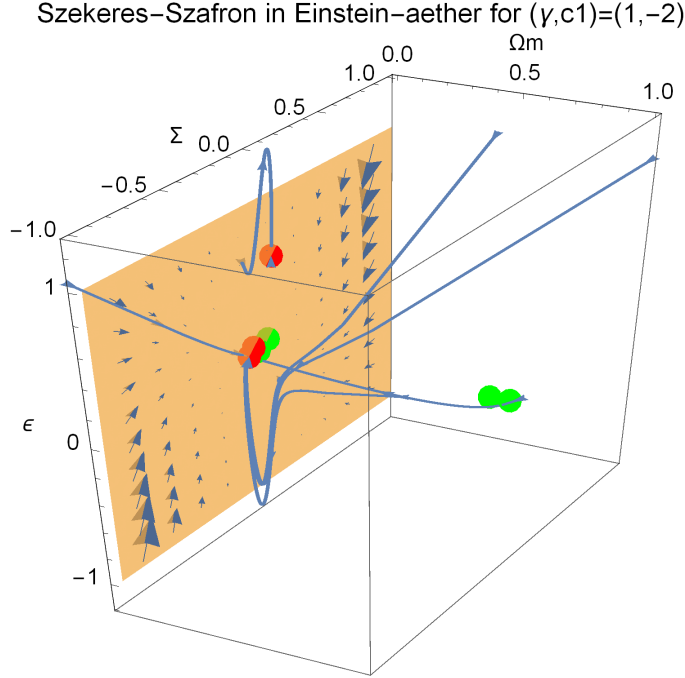}}\caption{Phase
portrait for the Szekeres-Szafron system in Einstein-\ae ther gravity, for
$\left(  \gamma,c_{1}\right)  =\left(  1,-2\right)  $. With red color are
marked the two Kasner attractors, points $B_{3}$ and $B_{6}$ while with green
color are marked the unstable critical points.}%
\label{fig5}%
\end{figure}

Let us demonstrate the results by considering the free parameters to be
$\left(  \gamma,c_{1}\right)  =\left(  1,-2\right)  $. In that case, the
dynamical system (\ref{sz.12})-(\ref{sz.14}) admits seven critical point, two
are stable and five are unstable points. The stable points are the $B_{3}$ and
the $B_{6}$ points. In Fig. \ref{fig5} we present the phase portrait for the
Einstein-\ae ther Szekeres-Szafron system for those specific values of the
free parameters. 

\section{Conclusions}

\label{sec5}

In this work, we performed a detailed study on exact solutions for
inhomogeneous spacetimes in the Einstein-\ae ther theory. Specifically we
studied the existence of exact solutions in Einstein-\ae ther gravity which
generalize the Szekeres solutions of GR.

For the \ae ther field we did the simplest selection by assuming that it is
the comoving observer. Indeed that it is not the general selection but it is
required if we assume the existence of a FLRW limit in the resulting
spacetimes \cite{roume}. For that specific selection of the \ae ther field,
the corresponding energy momentum tensor is calculated to be diagonal, hence
the constraint conditions provided by the field equations are those of GR.
Additionally, there is a new set of constraint conditions which follow by the
equation of motion for the \ae ther field.

The latter constraints provide conditions for the coefficient constants for
the \ae ther field, or constraints for the functional form of the scale
factors. From the line element of our consideration and for that specific
selection for the \ae ther field we found that the coefficient constants
$c_{3}$, $c_{4}$ for the \ae ther field, do not contribute in the dynamical
system. While when\footnote{In general, the algebraic constraint is
$c_{1}+c_{2}+c_{3}=0$, however, without loss of generality we selected
$c_{3}=0$.} $c_{1}+c_{2}=0$, is the unique case where inhomogeneous solutions exist.

The \ exact solutions we found describe spacetimes which belong to the two
classes of Szekeres, the inhomogeneous Kantowski-Sachs generalized spaces and
the inhomogeneous FLRW generalized space. However, the scale factors in this
case satisfy the modified field equations as given by the Einstein-\ae ther
theory for the Kantowski-Sachs and the FLRW spacetimes. On the other hand, for
arbitrary value of the coefficient constants $c_{1}$ and $c_{2}$, the unique
solution is that of a FRLW-like spacetime.

The stability of the solutions of the Szekeres spacetime in Einstein-aether
theory studied from where we find that the field equations evolve more
variously in Einstein-\ae ther than in GR; there are new critical points which
describe Kantowski-Sachs universes, while the stability of the critical points
with similar physical behaviour with that of GR change in a way to have as
attractors, Kasner universes, Kantowski-Sachs universes or spatially flat FLRW
universes with nonzero matter contribution in the universe. Contrary to GR the
attractors describe Kasner spacetimes.

This analysis contributes to the subject of existence of exact solutions in
Einstein-\ae ther theory.\ The novelty of this work is that we proved for the
first time in the literature the existence of inhomogeneous exact solutions in
the Einstein-\ae ther theory, by assuming extensions of the Szekeres
spacetimes. Recall that the latter spacetimes in general do not admit any symmetry.

In GR, Szekeres spacetimes can been seen as pertubative FLRW spaces
\cite{persz}, in a similar way it is consequence to consider a similar
analysis. In a future work, we plan to extend our analysis for a generic
\ae ther field as also a more generic form for the spacetime which extends the
Szekeres family. Last but not least, the Einstein-aether theory describes the
classical limit of Ho\~{r}ava gravity \cite{gg1}, which means that the
solutions we found correspond and also hold for the Ho\~{r}ava theory.

\begin{acknowledgments}
The author want to thank the anonymous referee for the valuable comments which
helped to improve the presentation of this work.
\end{acknowledgments}

\appendix

\section{Formulas and expressions}

\label{ap1}

In this Appendix we present expressions to which we have referred before.

Coordinates of point $B_{5}$:%

\begin{align*}
-\frac{2(\gamma+(\gamma+2)c_{1}-2)^{2}}{(\gamma-2)}\Omega_{m}\left(
B_{5}\right)   &  =3(\gamma-2)(5\gamma-4)+2(\gamma(6\gamma-13)+14)c_{1}%
^{2}+(52\gamma-48)c+\\
&  +\sqrt{3\gamma-2}(2c_{1}-1)\sqrt{%
\begin{array}
[c]{c}%
9(\gamma-2)^{2}(3\gamma-2)+4(\gamma(\gamma(3\gamma-10)+24)-16)c_{1}^{2}+\\
+4(\gamma-2)(\gamma(3\gamma+20)-16)c_{1}%
\end{array}
},
\end{align*}%
\begin{align*}
12(2c_{1}-1)(\gamma+(\gamma+2)c_{1}-2)\Sigma\left(  B_{5}\right)   &
=\gamma(3\gamma-8)-2(\gamma-2)(3\gamma-2)c_{1}+4+\\
&  -\sqrt{3\gamma-2}\sqrt{%
\begin{array}
[c]{c}%
9(\gamma-2)^{2}(3\gamma-2)+4(\gamma(\gamma(3\gamma-10)+24)-16)c_{1}^{2}+\\
+4(\gamma-2)(\gamma(3\gamma+20)-16)c_{1}%
\end{array}
},
\end{align*}

\begin{align*}
144(2c_{1}-1)(\gamma+(\gamma+2)c_{1}-2)^{2}\epsilon\left(  B_{5}\right)   &
=-\left(
\begin{array}
[c]{c}%
(8-3\gamma)\gamma+2(\gamma-2)(3\gamma-2)c_{1}-4+\\
\sqrt{3\gamma-2}\sqrt{%
\begin{array}
[c]{c}%
9(\gamma-2)^{2}(3\gamma-2)+\\
+4(\gamma(\gamma(3\gamma-10)+24)-16)c_{1}^{2}+4(\gamma-2)(\gamma
(3\gamma+20)-16)c_{1}%
\end{array}
}%
\end{array}
\right)  \times\\
&  \times\left(
\begin{array}
[c]{c}%
3(\gamma-4)\gamma-6\left(  \gamma^{2}-4\right)  c_{1}^{2}+2(3\gamma
(\gamma+2)-16))c_{1}+12+\\
\sqrt{3\gamma-2}\sqrt{%
\begin{array}
[c]{c}%
9(\gamma-2)^{2}(3\gamma-2)+\\
+4(\gamma(\gamma(3\gamma-10)+24)-16)c_{1}^{2}+4(\gamma-2)(\gamma
(3\gamma+20)-16))c_{1}%
\end{array}
}%
\end{array}
\right)  .
\end{align*}

\bigskip

Coordinates of point $B_{8}$:%
\begin{align*}
-\frac{2(\gamma+(\gamma+2)c_{1}-2)^{2}}{\left(  \gamma-2\right)  }\Omega
_{m}\left(  B_{8}\right)   &  =3(\gamma-2)(5\gamma-4)+2(\gamma(6\gamma
-13)+14)c_{1}^{2}+(52\gamma-48)c_{1}+\\
&  \sqrt{3\gamma-2}(1-2c_{1})\sqrt{%
\begin{array}
[c]{c}%
9(\gamma-2)^{2}(3\gamma-2)+4(\gamma(\gamma(3\gamma-10)+24)-16)c_{1}^{2}+\\
+4(\gamma-2)(\gamma(3\gamma+20)-16)c_{1}%
\end{array}
},
\end{align*}%
\begin{align*}
12(2c_{1}-1)(\gamma+(\gamma+2)c_{1}-2)\Sigma\left(  B_{5}\right)   &
=\gamma(3\gamma-8)-2(\gamma-2)(3\gamma-2)c_{1}+4+\\
&  \sqrt{3\gamma-2}\sqrt{%
\begin{array}
[c]{c}%
9(\gamma-2)^{2}(3\gamma-2)+4(\gamma(\gamma(3\gamma-10)+24)-16)c_{1}^{2}+\\
+4(\gamma-2)(\gamma(3\gamma+20)-16)c_{1}%
\end{array}
},
\end{align*}%
\begin{align*}
-144(2c_{1}-1)(\gamma+(\gamma+2)c_{1}-2)^{2}\epsilon\left(  B_{5}\right)   &
=\left(
\begin{array}
[c]{c}%
\gamma(3\gamma-8)-2(\gamma-2)(3\gamma-2)c_{1}+4+\\
\sqrt{3\gamma-2}\sqrt{%
\begin{array}
[c]{c}%
9(\gamma-2)^{2}(3\gamma-2)+4(\gamma(\gamma(3\gamma-10)+24)-16)c_{1}^{2}+\\
+4(\gamma-2)(\gamma(3\gamma+20)-16)c_{1}%
\end{array}
}%
\end{array}
\right)  \times\\
&  \times\left(
\begin{array}
[c]{c}%
-3(\gamma-4)\gamma-2(3\gamma(\gamma+2)-16)c_{1}-12+6\left(  \gamma
^{2}-4\right)  c_{1}^{2}\\
+\sqrt{3\gamma-2}\sqrt{%
\begin{array}
[c]{c}%
9(\gamma-2)^{2}(3\gamma-2)+4(\gamma-2)(\gamma(3\gamma+20)-16)c_{1}\\
+4(\gamma(\gamma(3\gamma-10)+24)-16)c_{1}^{2}%
\end{array}
}%
\end{array}
\right)  .
\end{align*}

\end{document}